\documentclass[12pt]{iopart}
\usepackage{iopams}
\usepackage{graphicx}
\begin{document}
\bibliographystyle{plain}
\title[On the recognition of fundamental physical
principles] {On the recognition of fundamental physical principles
in recent atmospheric-environmental studies}
\author{Gerhard Kramm$^1$, Ralph Dlugi$^2$ and Michael Zelger$^2$}
\address{$^1$ Geophysical Institute, University of Alaska Fairbanks, 903
Koyukuk Drive Fairbanks, AK 99775-7320, USA}
\address{$^2$ Arbeitsgruppe Atmosph\"arische Prozesse (AGAP),
Gernotstraße 11, D-80804 Munich, Germany}
\eads{\mailto{kramm@gi.alaska.edu}, \mailto{rdlugi@gmx.de},
\mailto{michael@zelger-net.de}}
\date{\today}
\begin{abstract}
In this paper, so-called alternative mass balance equations for
atmospheric constituents published recently are assessed in
comparison with the true local mass balance equations deduced from
exact integral formulations. It is shown that these "alternative"
expressions appreciably violate the physical law of the conservation
of mass as expressed by the equation of continuity. It is also shown
that terms of these "alternative" mass balance equations have
different physical units, a clear indication that these
"alternative" expressions are incorrect. Furthermore, it is argued
that in the case of "alternative" mass balance equations a real
basis for Monin-Obukhov similarity laws does not exist. These
similarity laws are customarily used to determine the turbulent
fluxes of momentum, sensible heat and matter in the so-called
atmospheric surface layer over even terrain. Moreover, based on
exact integral formulations a globally averaged mass balance
equation for trace species is derived. It is applied to discuss the
budget of carbon dioxide on the basis of the globally averaged
natural and anthropogenic emissions and the globally averaged uptake
caused by the terrestrial biosphere and the oceans.
\end{abstract}
\pacs{92.60.-e, 92.60.hg, 92.70.-j}
\maketitle
\setcounter{section}{0} \section{Introduction} \noindent Recently,
Finnigan et al. \cite{Fin1} and many others \cite{Aub, Fei, Fin2,
Fin3, Lee1, Lee2, Liu, Mas, Sog} proposed "alternative" forms of
balance equations of atmospheric constituents. Unfortunately, these
equations are not self-consistent and their terms are afflicted with
different physical units. The latter is a clear indication that
something is going wrong. Since mainly the results of turbulent
fluxes of carbon dioxide ($CO_2$) published recently were harmfully
affected by these "alternative" mass balance equations, these
results must be assessed carefully, especially in front of the
debate on global warming due to an increase of greenhouse
gases like $CO_2$.\\
\indent To assess these "alternative" mass balance equations it is
indispensable to compare them with the correct ones. Therefore, in
section 2 we will derive the correct local mass balance equations
from exact integral mass balance equations and discuss them
physically. It is shown in section 3 that one of the fundamental
physical principles namely the conservation of mass (as expressed by
the equation of continuity) is notably violated by the "alternative"
mass balance equations. Some aspects of Reynolds averaging are
briefly, but thoroughly discussed in section 4. In identifying an
adequate averaging time, we argue that the Allan variance criterion
\cite{All} combined with the wavelet analysis is much more favorable
than block averaging as suggested, for instance, by Finnigan et al.
\cite{Fin1} and Treviño and Andreas \cite{Tre}. In section 5, we
present the Reynolds averaged mass balance equations for a turbulent
fluid and compare it with that of the "alternative" forms. In
section 6, these different forms of mass balance equations are
vertically integrated by assuming horizontal homogeneity and the
results obtained are assessed. We show in section 7 that in the case
of the "alternative" form of the mass balance equations a real basis
for Monin-Obukhov similarity laws does not exist. Consequently, such
similarity laws customarily used for determining the flux densities
(hereafter, a flux density is simply called a flux) of long-lived
trace gases like $CO_2$ would be obsolete. Based on exact integral
formulations a globally averaged mass balance equation for trace
species is derived in section 8 to underline that the emission and
uptake of matter serve as boundary conditions. This result is in
complete contrast to the "alternative" mass balance equation in
which emission and uptake are considered as volume-related
source/sink functions. Final remarks and our conclusions are listed
in section 9.
\begin{figure}[t]
\begin{center}
\includegraphics[width=0.7\textwidth,height=!]{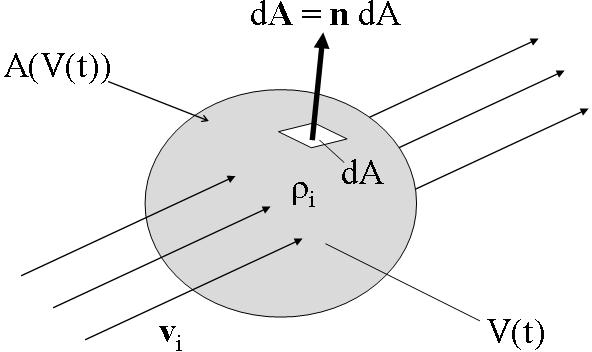}
\end{center}
\caption{Transfer of matter through the surface $A(V(t))$ of a given
volume $V(t)$.} \label{Figure1}
\end{figure}
\section{The mass balance equations of atmospheric constituents for a
macroscopic fluid} \noindent First, we will derive the conservation
laws for a macroscopic fluid (often called a molecular flow). The
total rate of change in the mass $M_i$ of a constituent, $i = 0, 1,
\dots, N$, within an arbitrary volume $V(t)$ that also depends on
time, $t$, can be expressed by \cite{Gro}
\begin{equation}\label{2.1}
\frac{dM_i}{dt} = \frac{d}{dt}\int_{V(t)} \rho_i\;dV\;,
\end{equation}
\noindent where $\rho_i = M_i/V$ is the corresponding partial
density. This quantity $dM_i/dt$ is equal to the material flow of
the $i^{th}$ component into and out of this volume through its
surface $A(V(t))$, i.e., the exchange of material between the volume
and its surroundings, and the net production/depletion $\sigma_i =
f(\rho_1, \dots, \rho_N)$ of the $i^{th}$ constituent per unit
volume owing to chemical reactions (and phase transition effects)
that occur inside $V$. Thus, we have (see Figure \ref{Figure1})
\begin{equation}\label{2.2}
\frac{d}{dt}\int_{V(t)} \rho_i\;dV = - \int_{A(V(t))} \rho_i\;
\textbf{v}_N \cdot d\textbf{A} + \int_{V(t)} \sigma_i\;dV\;.
\end{equation}
\noindent Here, $\textbf{v}_N = \textbf{v}_i - \textbf{v}$ is the
net velocity that characterizes the diffusion of the $i^{th}$
constituent with respect to the barycentric flow, considered as
being of Newtonian kind. The quantity  $\textbf{v}_i$ is the
individual velocity, and $\textbf{v}$ denotes the barycentric
velocity defined by
\begin{equation}\label{2.3}
\textbf{v} = \frac{1}{\rho} \sum_{i=0}^N \rho_i\;\textbf{v}_i\;,
\end{equation}
\noindent where
\begin{equation}\label{2.4}
\rho = \frac{M}{V} = \frac{1}{V} \sum_{i=0}^N M_i = \sum_{i=0}^N
\rho_i
\end{equation}
\noindent is the total density of air. The use of the barycentric
velocity as a reference velocity, of course, is not the only
possibility to describe diffusion. Prigogine \cite{Pri} deduced from
the entropy principle that in systems with mechanical equilibrium
diffusion processes can be related to an arbitrary reference
velocity. Herbert \cite{Her2, Her3} discussed the general
application of Prigogine's diffusion theorem to the atmosphere and
some specific invariance properties of the thermodynamic laws as
well as various alternative relations to describe Fick-type mass
diffusion in a (diluted) binary gas mixture such as the atmosphere.
Nevertheless, in our contribution the diffusion is related to the
barycentric velocity and the quantity $\textbf{J}_i = \rho_i
\textbf{v}_N = \rho_i (\textbf{v}_i - \textbf{v})$ is denoted as the
diffusion flux of the $i^{th}$ constituent. Furthermore,
$d\textbf{A} = \textbf{n}\;dA$ is a vector normal to the surface of
the volume $V$ with the unit vector $\textbf{n}$ and the magnitude
$dA$. The unit vector is counted positive from inside to outside of
the volume. Note that $i = 0$ denotes that portion of dry air which
is chemically inert, i.e., this portion does not contain any
chemically active atmospheric constituent. Furthermore, the
occurrence of the surface integral in Eq. (\ref{2.2}) means that we
consider a system that is open in the sense of thermodynamics, when
the exchange of energy is allowed, too. It is the most general one.\\
\indent Since the volume is considered as time-dependent, the
differentiation of the volume with respect to time is also required.
It can be performed by applying \emph{Leibnitz's integral theorem}.
In doing so, we obtain for the total rate of change, $dM_i/dt$,
\begin{equation}\label{2.5}
\frac{d}{dt}\int_{V(t)} \rho_i\;dV = \int_{V(t)}
\frac{\partial{\rho_i}}{\partial{t}}\;dV + \int_{A(V(t))} \rho_i\;
\textbf{v} \cdot d\textbf{A}\;.
\end{equation}
\noindent Thus, combining equations (\ref{2.2}) and (\ref{2.5})
results in
\begin{eqnarray}\label{2.6}
\nonumber \frac{d}{dt}\int_{V(t)} \rho_i\;dV &=& \int_{V(t)}
\frac{\partial{\rho_i}}{\partial{t}}\;dV + \int_{A(V(t))} \rho_i\;
\textbf{v} \cdot d\textbf{A}\\
&=& - \int_{A(V(t))} \textbf{J}_i \cdot d\textbf{A} + \int_{V(t)}
\sigma_i\;dV
\end{eqnarray}
\noindent or
\begin{equation}\label{2.7}
\int_{V(t)} \frac{\partial{\rho_i}}{\partial{t}}\;dV +
\int_{A(V(t))}\bigl(\rho_i \; \textbf{v} + \textbf{J}_i\bigr) \cdot
d\textbf{A} = \int_{V(t)} \sigma_i\;dV\;.
\end{equation}
\noindent This expression is called the \emph{integral balance
equation} of the $i^{th}$ atmospheric constituent. It is a very
general formulation and not affected by any kind of averaging
process usually applied in deriving local balance equations for
turbulent systems like the atmospheric boundary layer. McRae and
Russell \cite{Rae}, for instance, used such an integral formulation
for practical purposes by considering the domain of the South Coast
Air Basin of Southern California.\\
\indent By applying \emph{Gauss' integral theorem} (often called the
divergence theorem), the surface integral occurring in Eq.
(\ref{2.7}) can by replaced by a volume integral so that we obtain
\begin{equation}\label{2.8}
\int_{V(t)} \frac{\partial{\rho_i}}{\partial{t}}\;dV + \int_{V(t)}
\nabla \cdot \bigl(\rho_i \; \textbf{v} + \textbf{J}_i\bigr)\;dV =
\int_{V(t)} \sigma_i\;dV\;,
\end{equation}
\noindent where $\nabla$ is the nabla (or del) operator. Since in
Eq. (\ref{2.8}) the shape of the volume is arbitrary, we may also
write:
\begin{equation}\label{2.9}
\frac{\partial{\rho_i}}{\partial{t}} + \nabla \cdot \bigl(\rho_i \;
\textbf{v} + \textbf{J}_i\bigr)=  \sigma_i\;.
\end{equation}
\noindent This equation is called the \emph{local balance equation}
of the $i^{th}$ constituent \cite{Bus, Gro, Kra00, Rau, Sei, Stu}.
The term $\partial{\rho_i}/\partial{t}$ is the local temporal change
of $\rho_i$, and the quantities $\rho_i \textbf{v}$ and
$\textbf{J}_i$ are frequently called the \emph{convective} and
\emph{non-convective transports} of matter, respectively. It is
obvious that in the case of a closed or isolated thermodynamic
system the divergence term vanishes because it only describes the
exchange of matter between the system and its surroundings. As the
use of Gauss' integral theorem requires that some mathematical
pre-requisites have to be fulfilled, the local balance equation
(\ref{2.9}) is somewhat lesser valid than the integral balance
equation (\ref{2.7}). Nevertheless, any integration of Eq.
(\ref{2.9}) over a certain volume as performed, for instance, by
numerical models of the troposphere must be in agreement with Eq.
(\ref{2.7}).\\
\indent Summing Eq. (\ref{2.9}) over all substances, $i = 0, 1.
\dots, N$, provides
\begin{equation}\label{2.10}
\sum_{i=0}^N \frac{\partial{\rho_i}}{\partial{t}} + \sum_{i=0}^N
\nabla \cdot \bigl(\rho_i \; \textbf{v} + \textbf{J}_i\bigr)=
\sum_{i=0}^N \sigma_i\;.
\end{equation}
\noindent Since, according to de Lavoisier's law, mass is conserved
in chemical reactions (and/or phase transition processes; i.e., the
transformation of mass into energy as expressed by Einstein's
formula is unimportant in the case of atmospheric trace
constituents), we have
\begin{equation}\label{2.11}
\sum_{i=0}^N \sigma_i = 0\;.
\end{equation}
\noindent Furthermore, with the aid of Eqs. (\ref{2.3}) and
(\ref{2.4}) and the definition of the diffusion flux, we obtain
\begin{equation}\label{2.12}
\sum_{i=0}^N \frac{\partial{\rho_i}}{\partial{t}} =
\frac{\partial{}}{\partial{t}} \Bigl(\sum_{i=0}^N \rho_i\Bigr) =
\frac{\partial{\rho}}{\partial{t}}
\end{equation}
\noindent and
\begin{eqnarray}\label{2.13}
\nonumber \sum_{i=0}^N \nabla \cdot (\rho_i \; \textbf{v} +
\textbf{J}_i) &=& \nabla \cdot \Bigl(\sum_{i=0}^N \bigl(\rho_i \;
\textbf{v} + \rho_i(\textbf{v}_i - \textbf{v})\bigr)\Bigr) \\
&=& \nabla \cdot \Bigl(\sum_{i=0}^N \bigl(\rho_i \;
\textbf{v}_i\bigr)\Bigr) = \nabla \cdot \bigl(\rho\;
\textbf{v}\bigr) \;.
\end{eqnarray}
\noindent Thus, Eq, (\ref{2.10}) leads to the macroscopic balance
equation for the total mass per unit volume \cite{Gro, Gar, Kra00,
Lan1, Rau, Stu}
\begin{equation}\label{2.14}
\frac{\partial{\rho}}{\partial{t}} + \nabla \cdot \bigl(\rho \;
\textbf{v}\bigr)= 0\;.
\end{equation}
\noindent It represents the formulation of the conservation of mass
on the local scale and is customarily called the \emph{equation of
continuity}.\\
\indent If we use the mass fraction $c_i = \rho_i/\rho$, then Eq.
(\ref{2.9}) will read
\begin{equation}\label{2.15}
\frac{\partial{\bigl(\rho\; c_i\bigr)}}{\partial{t}} + \nabla \cdot
\bigl(\rho\; c_i \; \textbf{v} + \textbf{J}_i\bigr)=  \sigma_i\;.
\end{equation}
\noindent Applying the product rule of differentiation to this
equation yields
\begin{equation}\label{2.16}
\rho\;\frac{\partial{c_i}}{\partial{t}} + c_i
\Bigl(\underbrace{\frac{\partial{\rho}}{\partial{t}} + \nabla \cdot
\bigl(\rho \; \textbf{v}\bigr)}_{=\;0\; (see Eq.
(\ref{2.14}))}\Bigr) + \rho\;\textbf{v} \cdot \nabla{c_i} + \nabla
\cdot \textbf{J}_i = \sigma_i
\end{equation}
\noindent or
\begin{equation}\label{2.17}
\rho\Bigl(\frac{\partial{c_i}}{\partial{t}} + \textbf{v} \cdot
\nabla{c_i}\Bigr) + \nabla \cdot \textbf{J}_i = \sigma_i\;,
\end{equation}
\noindent where
\begin{equation}\label{2.18}
\frac{dc_i}{dt} = \frac{\partial{c_i}}{\partial{t}} + \textbf{v}
\cdot \nabla{c_i}
\end{equation}
\noindent is the substantial or total derivative with respect to
time. Equation (\ref{2.17}) may be called the
\emph{advection-diffusion equation} of a chemically active
atmospheric constituent.
\section{The "alternative" mass balance equations of atmospheric
constituents} \noindent Recently, several authors proposed a mass
balance as an "alternative" form to the conservation equation
(\ref{2.9}). This "alternative" form reads \cite{Aub, Fei, Fin1,
Fin2, Fin3, Lee1, Lee2, Liu, Mas, Sog}
\begin{equation}\label{3.1}
\frac{\partial{\rho_i}}{\partial{t}} + \nabla \cdot \bigl(\rho_i \;
\textbf{v}\bigr)= S_i\;.
\end{equation}
\noindent In this equation, the diffusion flux $\textbf{J}_i$ is
generally ignored. This flux, however, is not negligible because, as
expressed in Eq. (\ref{2.9}), it must represent the local balance
equation for a macroscopic fluid. If $\textbf{J}_i$ would not occur,
diffusion of the $i^{th}$ constituent in a macroscopic fluid could
not be quantified. This means, for instance, that the sedimentation
of airborne particles like aerosol and ice particles and water drops
would generally be excluded. Moreover, this "alternative" form is
applied to quantify the sinks or sources of $CO_2$ inside canopies
of tall vegetation like forests by a method where the term $S_i$ is
called a source/sink inside the volume $V$ of the fluid. However,
considering the derivation of Eq. (\ref{2.9}), an uptake or emission
of a substance by plant elements or soil is a process at the
boundary of a fluid and, therefore, has to be described in terms of
boundary conditions, as substantiated by the integral balance
equation (\ref{2.7}). Therefore, $S_i$ is not a biological
source/sink term. Such a term must not occur in the local form of a
balance equation. Boundary conditions only occur when the local mass
balance equation is integrated, in complete agreement with Eq.
(\ref{2.7}). The "alternative" equation (\ref{3.1}), however, cannot
be deduced from any integral mass balance equation. Finnigan et al.
\cite{Fin1} introduced it into the literature in an unforced manner
and without any physical justification. One of the physical
consequences related to this "alternative" mass balance equation is
that the biological source/sink would explicitly cause a temporal
change in the partial density. Whereas Eq. (\ref{2.9}) clearly
substantiates that only the divergence of the convective and
non-convective fluxes contributes to a temporal change of the
partial density.\\
\indent Finnigan et al. \cite{Fin1} and many others \cite{Aub, Fei,
Fin2, Fin3, Lee1, Lee2, Liu, Mas, Sog} expressed $S_i$ as a surface
source (instead of a volume-related source or sink due to chemical
reactions or/and phase transition processes) by
\begin{equation}\label{3.2}
S_i = S_i\bigl(\textbf{x}\bigr)\;\delta \bigl(\textbf{x} -
\textbf{x}_0\bigr)\;,
\end{equation}
\noindent where $\delta(\textbf{x} - \textbf{x}_0)$ is Dirac's delta
function (from a mathematical point of view a distribution).
Finnigan et al. \cite{Fin1} argued that "the source term,
$S_i(\textbf{x})$, is multiplied by the Dirac delta function,
signifying that the source is zero except on the ground and
vegetation surfaces, whose locus is $\textbf{x}_0$". This is
incorrect at least by two reasons: As mentioned before, the physical
meaning of $S_i$ in Eq.(\ref{3.1}) must be a source or sink within
the volume, but not at the boundaries, which are represented by
boundary conditions. Furthermore, the Dirac delta function has the
property that $\delta(\textbf{x} - \textbf{x}_0) = 0$ for all
$\textbf{x} \neq \textbf{x}_0$ \cite{Bra, Dir, Lan2, Mes, Ril},
i.e., there is only the infinitesimal region (namely when
$\textbf{x} = \textbf{x}_0$) at which a source or sink of matter is
defined. In addition, in various textbooks \cite{Lan2, Mes}, it is
pointed out that $\delta(\textbf{x} - \textbf{x}_0) = \infty$ for
$\textbf{x} = \textbf{x}_0$. Thus, Eq. (\ref{3.2}) and, hence, Eq.
(\ref{3.1}) would become meaningless for $\textbf{x} =
\textbf{x}_0$. As pointed out later on, the main properties of the
$\delta$-function are defined by the integral over a region
containing $\textbf{x} = \textbf{x}_0$.\\
\indent Uptake and emission of a constituent by plants, soil and/or
water systems are surface effects, expressed, for instance, in SI
units by $kg/(m^2 s)$; whereas the local temporal change in Eqs.
(\ref{2.9}) and (\ref{2.15}) has the SI units $kg/(m^3 s)$. If we
express Dirac's delta function using SI units we will have $m^{-3}$.
Therefore, the term
$S_i(\textbf{x})\;\delta(\textbf{x}-\textbf{x}_0)$ would have the SI
units $kg/(m^5 s)$. It is indispensable that in any physical
equation its terms must have identical physical units. This
fundamental requirement, however, is not satisfied in the case of
the "alternative" forms of mass balance equations. There is evidence
(e.g., Eq. (6) of Finnigan et al. \cite{Fin1}, Eq. (11) of Finnigan
\cite{Fin2}, Eqs. (2.2) and (3.8) of Lee et al. \cite{Lee2}, and Eq.
(14) of Massman and Tuovinen \cite{Mas}) that the biological
source/sink term is, indeed, expressed by the units of a flux of
matter, i.e., in $kg/(m^2 s)$, even though, as expressed by the
local temporal change of the partial density
($\partial{\rho_i}/\partial{t}$), $kg/(m^3 s)$ is
unequivocally required.\\
\indent Nevertheless, assuming for a moment that Eq. (\ref{3.2}) is
reasonable, then Eq. (\ref{2.11}) would result in
\begin{equation}\label{3.3}
\sum_{i=1}^N S_i = \sum_{i=1}^N S_i\bigl(\textbf{x}\bigr)\;\delta
\bigl(\textbf{x} - \textbf{x}_0\bigr)\neq 0\;,
\end{equation}
\noindent and, hence, Eq. (\ref{2.10}) in
\begin{equation}\label{3.4}
\frac{\partial{\rho}}{\partial{t}} + \nabla \cdot \bigl(\rho \;
\textbf{v}\bigr) = \sum_{i=1}^N S_i\bigl(\textbf{x}\bigr)\;\delta
\bigl(\textbf{x} - \textbf{x}_0\bigr)\neq 0\;.
\end{equation}
\noindent The right-hand side of this equation is not equal to zero
because atmospheric constituents have different natural and
anthropogenic origins and different sinks. This means: if Eq.
(\ref{3.2}) would be reasonable, than one of the fundamental laws of
physics, namely the conservation of the total mass on a local scale,
as expressed by the equation of continuity (\ref{2.14}), is notably
violated by the "alternative" forms of mass balance equations.
\section{Reynolds averaging calculus}
\noindent Conservation laws for a turbulent fluid can be derived by
using Reynolds' averaging calculus, i.e., decomposition of any
instantaneous field quantity $\varphi(\textbf{r})$ like
$\rho_i(\textbf{r})$ and $\textbf{v}(\textbf{r})$ by
$\varphi(\textbf{r}) = \overline{\varphi} + \varphi'$ and subsequent
averaging according to \cite{Her1, Kra00, Mie}
\begin{equation}\label{4.1}
\overline{\varphi} = \overline{\varphi(\textbf{r})} = \frac{1}{G}
\int_{G} \varphi \bigl(\textbf{r}, \textbf{r}'\bigr)\;dG'\;,
\end{equation}
\noindent where $\overline{\varphi}$ is the average in space and
time of $\varphi(\textbf{r})$, and the fluctuation $\varphi'$ is the
difference between the former and the latter. Here, $\textbf{r}$ is
the four-dimensional vector of space and time in the original
coordinate system, $\textbf{r}'$ is that of the averaging domain $G$
where its origin, $\textbf{r}' = 0$, is assumed to be $\textbf{r}$,
and $dG' = d^3r\;dt'$. The averaging domain $G$ is given by $G =
\int_{G} dG'$. Hence, the quantity $\overline{\varphi}$ represents
the mean values for the averaging domain $G$ at the location
$\textbf{r}$. Since $\overline{\overline{\varphi}} =
\overline{\varphi}$, averaging the quantity $\varphi(\textbf{r}) =
\overline{\varphi} + \varphi'$ provides $\overline{\varphi'} = 0$.
Any kind of averaging must be in agreement with these basic
definitions. These basic definitions cannot be undermined by
imperfect averaging procedures. Since, in accord with the ergodic
theorem, assemble averaging as expressed by Eq. (\ref{4.1}) may be
replaced by time averaging \cite{Lie}, the common practice, it is
indispensable to ask whether a time averaging procedure is in
complete agreement with the statistical description of turbulence or
not.\\
\indent A basic requirement for using a time averaging procedure is
that turbulence is statistically steady \cite{Fal, Lum, Ten}.
Consequently, sophisticated procedures for identifying
non-stationary effects (trends) are indispensable to prevent that
computed turbulent fluxes of atmospheric constituents are notably
affected by non-stationary effects. In order to identify such
stationary states in the off-line time series analysis, Werle et al.
\cite{Wer1} suggested the Allan-variance criterion \cite{All} as a
suitable and efficient tool. As argued by Percival and Guttorp
\cite{Per}, this variance is an appropriate measure for studying
long-memory processes because it can be estimated without bias and
with good efficiency for such processes, and it may be interpreted
as the Haar wavelet coefficient variance \cite{Fla, Kra99}. Percival
and Guttorp \cite{Per} generalized this to other wavelets
\cite{Kum}. Following various authors \cite{Bru, Hag, How, Katu},
wavelet decomposition seems to be well appropriate to study
turbulence data. Thus, the Allan variance criterion combined with
the wavelet analysis is much more favorable than simple block
averaging as suggested, for instance, by Finnigan et
al. \cite{Fin1} and Treviño and Andreas \cite{Tre}.\\
\begin{figure}[t]
\begin{center}
\includegraphics[width=0.8\textwidth,height=!]{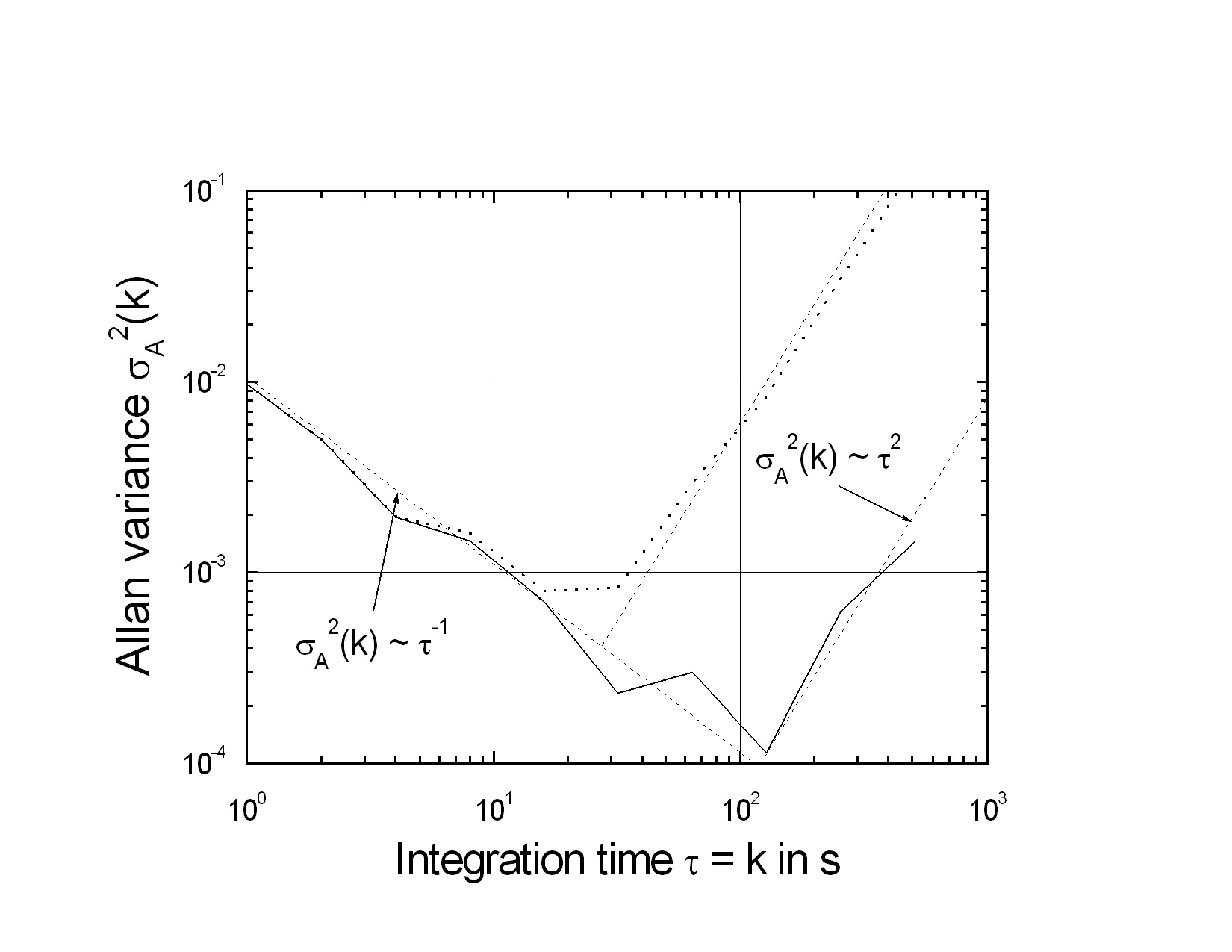}
\end{center}
\caption{Allan plots deduced from the time series data illustrated
in Figure \ref{Figure3} (solid line) and a data set which differs
from that by a ten times stronger drift (dotted line). The dashed
lines indicate white noise ($\tau^{- 1}$) and linear drift
($\tau^2$) behavior (adopted from \cite{Kra99}).} \label{Figure2}
\end{figure}
\begin{figure}[t]
\begin{center}
\includegraphics[width=0.8\textwidth,height=!]{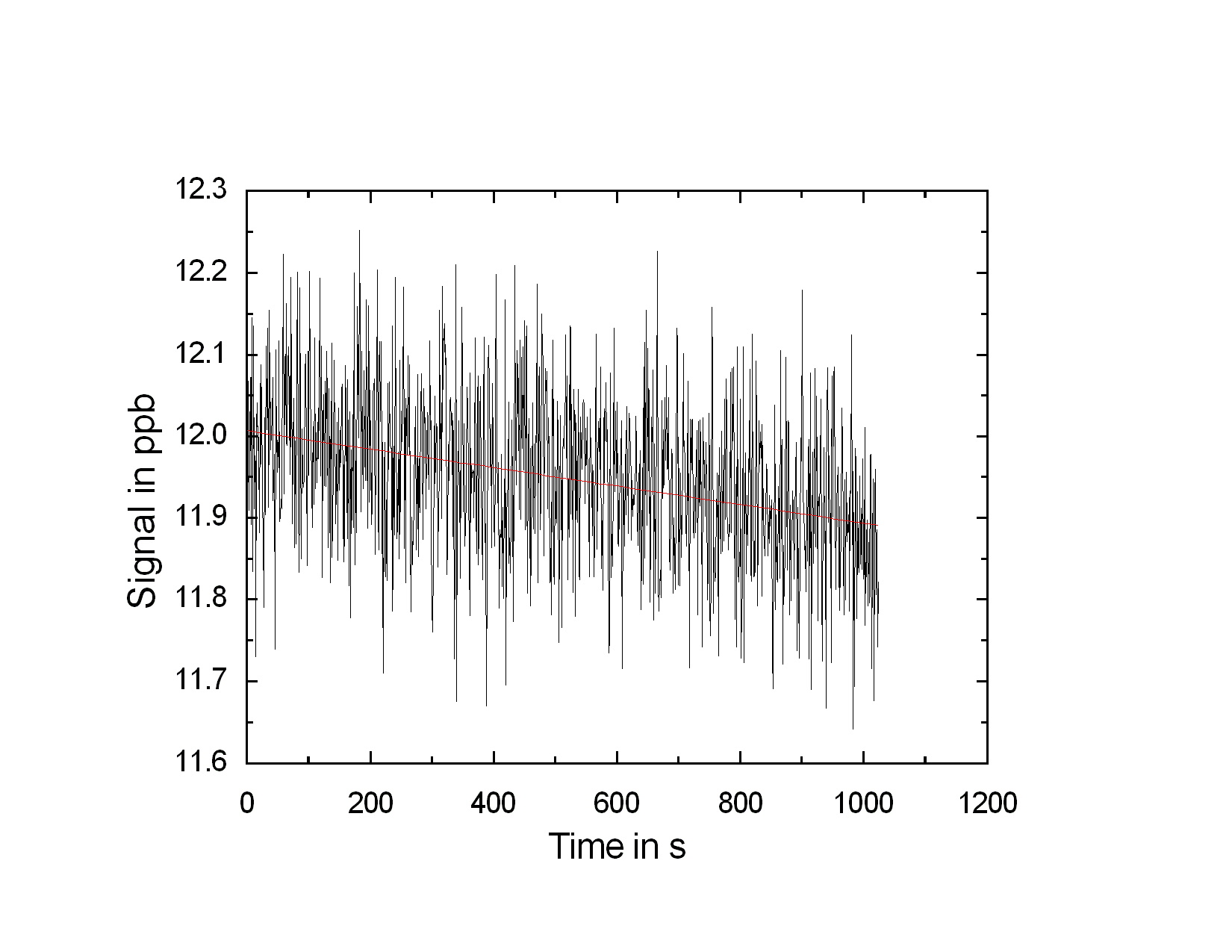}
\end{center}
\caption{Generated time series data of a concentration consisting of
an offset, a linear drift and a Gaussian distributed white noise.
The sample size amounts to $N = 2^{10} = 1024$ (adopted from
\cite{Wer2}).} \label{Figure3}
\end{figure}
\indent Figure \ref{Figure2} illustrates two Allan plots deduced
from a synthetic data set generated by the equation $Y_t = a + b\;t
+ G(\sigma)$ (see Figure \ref{Figure3}), where $a$ is an offset,
$b\; t$ is a linear drift and $G(\sigma)$ is a Gaussian distributed
white noise (lower trace) and a data set containing a ten times
stronger drift (upper trace). In both cases $\sigma_A^2(k)$ was
calculated using Haar wavelet coefficients \cite{Kra99}. For
convenience, it is assumed that the sampling interval between
consecutive observations is constant and amounts to $\Delta{t}=
1\;s$. Thus, the time $\tau = k\;\Delta{t}$ is equivalent to the
averaging or integration time. As shown in Figure \ref{Figure3}, the
lower trace indicates a minimum of $\sigma_A^2(k)$ at an integration
time, denoted here as optimum integration time $\tau_{opt}$, of
about $130\; s$. Whereas the upper trace for the data set with the
ten times stronger drift suggests a $\tau_{opt}$ value of about
$25\;s$. For $\tau < \tau_{opt}$, the Allan plots show a $\tau^{-
1}$ behavior that is typical for white noise. Beyond $\tau_{opt}$,
the Allan plots obey the $\tau^2$ law which indicates that a linear
drift becomes dominant. Consequently, for $\tau > \tau_{opt}$,
stationary conditions as required by time averaging are not assured.
From this point of view, $\tau_{opt}$ may be considered as the
maximum averaging time \cite{Kra99}.
\section{The balance equations of
atmospheric constituents for a turbulent fluid} \noindent Applying
the conventional Reynolds' averaging calculus to Eq. (\ref{2.9})
yields \cite{Bus, Kra00}
\begin{equation}\label{5.1}
\frac{\partial{\overline{\rho_i}}}{\partial{t}} + \nabla \cdot
\Bigl(\overline{\rho_i}\;\overline{\textbf{v}} +
\overline{\rho_i'\;\textbf{v}'} + \overline{\textbf{J}_i}\Bigr)=
\overline{\sigma_i}\;.
\end{equation}
\noindent Here, the overbar represents the conventional Reynolds
average; whereas the prime denotes the departure from that. Equation
(\ref{5.1}) is called the balance equation for the $1^{st}$ moment
(or $1^{st}$-order balance equation). Here,
$\overline{\rho_i'\;\textbf{v}'}$  is the turbulent (or eddy) flux.
It is a non-convective flux, too.\\
\indent As mentioned before, the diffusion flux represents not only
molecular fluxes, but also the sedimentation of aerosol particles
affected by the gravity field. Thus, $\overline{\textbf{J}_i}$
cannot generally be neglected in comparison with the corresponding
second moment $\overline{\rho_i'\;\textbf{v}'}$. Even in the
immediate vicinity of the earth's surface it must not be ignored
because the wind vector vanishes at any rigid surface, and,
consequently, no exhalation or deposition of matter would be
possible. In the case of gaseous constituents
$\overline{\textbf{J}_i}$ becomes negligible in comparison with
$\overline{\rho_i'\;\textbf{v}'}$ when a fully turbulent flow is
considered. Obviously, Eq. (\ref{5.1}) also fulfils the conditions
$\sum_{i=0}^N \overline{\textbf{J}_i} = 0$ and $\sum_{i=1}^N
\overline{\sigma_i} = 0$, and, as required by the equation of
continuity in its averaged form,
\begin{equation}\label{5.2}
\frac{\partial{\overline{\rho}}}{\partial{t}} + \nabla \cdot
\Bigl(\overline{\rho}\;\overline{\textbf{v}} +
\overline{\rho'\;\textbf{v}'}\Bigr)= 0\;,
\end{equation}
\noindent the conditions $\sum_{i=0}^N
\overline{\rho_i}\;\overline{\textbf{v}}=
\overline{\rho}\;\overline{\textbf{v}}$ and $\sum_{i=0}^N
\overline{\rho_i'\;\textbf{v}'} = \overline{\rho'\;\textbf{v}'}$ are
fulfilled, too.\\
\indent If mass fractions are considered (see Eqs. (\ref{2.15}) to
(\ref{2.18})), then density-weighted averaging techniques \cite{Dut,
Her1, Kra95, Pic, Mie} should be applied in formulating local
balance equations for turbulent flows to avoid any simplification
\cite{Kra00, Kra06}.\\
\indent If we rearrange Eq. (\ref{5.1}) by using the equation of
continuity (\ref{5.2}) (it is similar to the rearrangement of Eq.
(\ref{2.15}) that leads to Eq. (\ref{2.17})) and ignoring the
covariance terms $\overline{\rho'\;\textbf{v}'}$ and
$\overline{\rho'\;c_i'}$, as customarily accepted within the
framework of the Boussinesq approximation, we will obtain
\begin{equation}\label{5.3}
\overline{\rho} \Bigl(\frac{\partial{\overline{c_i}}}{\partial{t}} +
\overline{\textbf{v}} \cdot \nabla{\overline{c_i}}\Bigr) + \nabla
\cdot \Bigl(\overline{\rho_i'\;\textbf{v}'} +
\overline{\textbf{J}_i}\Bigr)= \overline{\sigma_i}\;.
\end{equation}
\noindent It may be called the advection-diffusion equation of a
chemically active atmospheric constituent for a turbulent fluid.
Here, $\sigma_i$ still represents the gain or lost of matter due to
chemical reactions (and/or phase transition processes).\\
\indent For the purpose of comparison: Finnigan's \cite{Fin2} Eqs.
(8) as well as Eq. (3.7) and (10.8) of Lee et al. \cite{Lee2} read
\begin{equation}\label{5.4}
\frac{\partial{\overline{\rho_i}}}{\partial{t}} +
\overline{\textbf{v}} \cdot \nabla{\overline{\rho_i}} + \nabla \cdot
\Bigl(\overline{\rho_i'\;\textbf{v}'}\Bigr)= \overline{S_i}\;,
\end{equation}
\noindent where $S_i$ is still given by $S_i(\textbf{x}) \delta
(\textbf{x} - \textbf{x}_0)$ (see our Eq. (\ref{3.2})). Obviously,
Eq. (\ref{5.4}) is not self-consistent. First, as shown in Eq.
(\ref{5.3}), the mass fraction $c_i$ must occur in the two terms on
the left-hand side of Eq. (\ref{5.4}), but not the partial density
$\rho_i$. Second, the rearrangement of Eq. (\ref{2.15}) that leads
to Eq. (\ref{2.17}) requires that the right-hand side of Eq.
(\ref{3.4}) is always equal to zero so that Eq. (\ref{3.4}) equals
the equation of continuity (\ref{2.14}).
\section{Vertically integrated mass balance equations}
\noindent If we assume horizontally homogeneous conditions and
recognize that in such a case the vertical component of the mean
wind vector becomes nearly equal to zero ($\overline{w} \cong 0$),
Eq. (\ref{5.1}) will provide
\begin{equation}\label{6.1}
\frac{\partial{\overline{\rho_i}}}{\partial{t}} +
\frac{\partial}{\partial{z}}\Bigl(\overline{\rho_i'\;w'} +
\overline{J_{i,z}}\Bigr)= \overline{\sigma_i}\;,
\end{equation}
\noindent where, $w$ is the vertical component of the wind vector,
$J_{i,z}$ is the vertical component of the diffusion flux, and $z$
is the height above ground. Note that Archimedean effects are
related to the gravity field. The same is true in the case of the
hydrostatic pressure. Consequently, the vertical direction is
related to the gravity field\cite{Sun1}; but not to the normal
vector of a slope or a streamline as recently described, for
instance, by Finnigan et al. \cite{Fin1} and Finnigan \cite{Fin2}.
Especially over complex terrain trajectories and streamlines are
usually vary with time and position. Therefore, it is difficult to
find, for instance, a common reference streamline coordinate frame
for using sonic anemometry at different heights to estimate the
variation of turbulent fluxes of momentum, sensible heat and matter
with height. Without knowing a reference coordinate frame that is
invariant in space and time, the convergence/divergence of any trace
gases cannot be calculated \cite{Sun1, Sun2}. In contrast to a
streamline coordinate frame, one may use a terrain-following
coordinate frame as usually considered within the framework of
mesoscale meteorological modeling, but an exact transformation of
the governing equations is still indispensable. Fortunately, this
transformation is well known since more than three decades
(see, e.g., \cite{Piel}).\\
\indent The integration of this equation from the earth's surface
($z = 0$) to a certain height above ground ($z = h$),where a fully
turbulent flow is established, yields then
\begin{equation}\label{6.2}
\int_{0}^{h} \Bigl(\frac{\partial{\overline{\rho_i}}}{\partial{t}} -
\overline{\sigma_i}\Bigr) dz = \Bigl[\overline{\rho_i'\;w'} +
\overline{J_{i,z}}\Bigr]_{z=0} - \Bigl[\overline{\rho_i'\;w'} +
\overline{J_{i,z}}\Bigr]_{z=h}\quad.
\end{equation}
\noindent Assuming long-lived trace species ($\overline{\sigma_i}
\cong 0$) and stationary condition
($\partial{\overline{\rho_i}}/\partial{t}= 0$)leads to the constant
flux approximation expressed by
\begin{equation}\label{6.3}
\Bigl[\overline{\rho_i'\;w'} + \overline{J_{i,z}}\Bigr]_{z=h} =
\Bigl[\overline{\rho_i'\;w'} + \overline{J_{i,z}}\Bigr]_{z=0}\;.
\end{equation}
\noindent At the height $z = h$ the vertical component of the
diffusion flux of a trace gas can usually be ignored in comparison
with the vertical component of the corresponding eddy flux component
($|\overline{J_{i,z}}|_{z=h} \ll |\overline{\rho_i'\;w'}|_{z=h}$).
As already mentioned, the opposite is true in the immediate vicinity
of the earth's surface ($|\overline{J_{i,z}}|_{z=0} \gg
|\overline{\rho_i'\;w'}|_{z=0}$). Thus, Eqs. (\ref{6.2}) and
(\ref{6.3}) may be written as
\begin{equation}\label{6.4}
\int_{0}^{h} \Bigl(\frac{\partial{\overline{\rho_i}}}{\partial{t}} -
\overline{\sigma_i}\Bigr) dz \cong
\Bigl[\overline{J_{i,z}}\Bigr]_{z=0} -
\Bigl[\overline{\rho_i'\;w'}\Bigr]_{z=h}
\end{equation}
\noindent and
\begin{equation}\label{6.5}
\Bigl[\overline{\rho_i'\;w'}\Bigr]_{z=h} \cong \Bigl[
\overline{J_{i,z}}\Bigr]_{z=0}\;.
\end{equation}
\noindent In contrast to this, Eq. (\ref{5.4}) with  $S_i = S_i(z)\;
\delta(z)$ (see Eq. (\ref{3.2})) leads to \cite{Fin1, Fin2, Fin3,
Mon}
\begin{equation}\label{6.6}
\frac{\partial{\overline{\rho_i}}}{\partial{t}} +
\frac{\partial}{\partial{z}}\Bigl(\overline{\rho_i'\;w'}\Bigr)=
\overline{S_i(z)}\;\delta(z)\;.
\end{equation}
\noindent Here, an important inconsistency exists because
$\textbf{x}$ and $\textbf{x}_0$ are vectors so that
$\delta(\textbf{x} - \textbf{x}_0) = 0$ has to be expressed, for
instance, in Cartesian coordinates by $\delta(x - x_0)\delta(y -
y_0)\delta(z) = 0$ when $z_0 = 0$ is considered. Therefore, we would
have
\begin{equation}\label{6.7}
\frac{\partial{\overline{\rho_i}}}{\partial{t}} +
\frac{\partial}{\partial{z}}\Bigl(\overline{\rho_i'\;w'}\Bigr)=
\overline{S_i(z)}\;\delta \bigl(x - x_0\bigr)\;\delta \bigl(y -
y_0\bigr)\;\delta\bigl(z\bigr)\;.
\end{equation}
\noindent Nevertheless, following for a moment Finnigan and
disciples, the integration of this equation should yield \cite{Fin1,
Fin2, Fin3, Mon}
\begin{equation}\label{6.8}
\int_{0}^{h} \frac{\partial{\overline{\rho_i}}}{\partial{t}} dz = -
\Bigl[\overline{\rho_i'\;w'}\Bigr]_{z=h} + \overline{S_i}\;,
\end{equation}
\noindent where the biological source/sink term $\overline{S_i}$ is
given by
\begin{equation}\label{6.9}
\overline{S_i} = \int_0^h \overline{S_i(z)}\;\delta
\bigl(z\bigr)\;dz\;.
\end{equation}
\noindent Assuming steady-state condition yields then \cite{Mon}\\
\begin{equation}\label{6.10}
\Bigl[\overline{\rho_i'\;w'}\Bigr]_{z=h} = \overline{S_i}\;.
\end{equation}
\noindent Obviously, Eq (\ref{6.10}) looks similar like Eq.
(\ref{6.5}). However, from a mathematical point of view, Eq.
(\ref{6.9}) is faulty because Dirac's delta function has the
fundamental property that \cite{Bra, Dir, Lan2, Mes, Ril}
\begin{equation}\label{6.11}
\int_{-\infty}^{\infty} f\bigl(x\bigr)\;\delta \bigl(x - a\bigr)\;dx
= f\bigl(a\bigr)\;,
\end{equation}
\noindent where $f(x)$ is any function continuous at the point $x =
a$, and, in fact,
\begin{equation}\label{6.12}
\int_{a-\epsilon}^{a+\epsilon} f\bigl(x\bigr)\;\delta \bigl(x -
a\bigr)\; dx = f(a)\quad \textrm{for} \;\epsilon > 0
\end{equation}
\noindent or, with $b_1 = a-\epsilon < a$ and $b_2 = a+\epsilon >
a$,
\begin{equation}\label{6.13}
\int_{b_1}^{b_2} f\bigl(x\bigr)\;\delta \bigl(x - a\bigr)\; dx =
f\bigl(a\bigr)\quad \textrm{for} \;b_1 < a < b_2\quad.
\end{equation}
\noindent In the special case of $f(x) = 1$ this equation directly
provides
\begin{equation}\label{6.14}
\int_{b_1}^{b_2} \delta \bigl(x - a\bigr)\; dx = 1\;.
\end{equation}
\noindent This means that the range of integration must include the
point $x = a$, as expressed in Eq. (\ref{6.13}) by $b_1 < a < b_2$;
otherwise, the integral equals zero. Equation (\ref{6.9}) does not
fulfill this requirement. Consequently, we have
\begin{equation}\label{6.15}
\overline{S_i} = \int_0^h \overline{S_i(z)}\;\delta
\bigl(z\bigr)\;dz = 0\;.
\end{equation}
\noindent This result is generally valid for any value of
$z\;\geq\;0$.
\section{The bases for Monin-Obukhov similarity laws}
\noindent Similarity hypotheses according to Monin and Obukhov
\cite{Bar, Monin} are based on the pre-requisite that the turbulent
fluxes of momentum, sensible heat and matter are invariant with
height across the atmospheric surface layer (ASL). These similarity
hypotheses can be expressed by \cite{Monin, Busch, Pano84, Kra89}
\begin{equation}\label{7.1} \frac{\kappa\;z}{u_*} \frac{dU}{dz} =
\Phi_m \bigl(\zeta\bigr)\;,
\end{equation}
\begin{equation}\label{7.2}
\frac{\kappa\;z}{\Theta_*} \frac{d\overline{\Theta}}{dz} = \Phi_h
\bigl(\zeta\bigr)\;,
\end{equation}
\noindent and
\begin{equation}\label{7.3}
\frac{\kappa\;z}{c_{i,*}} \frac{d\overline{c_i}}{dz} = \Phi_i
\bigl(\zeta\bigr)\;.
\end{equation}
\noindent Here, $\kappa$ is the von Kármán constant, $u_*$ is the
friction velocity,  $U = |\overline{\textbf{v}_H}|$ is the magnitude
of the mean horizontal wind component, $\Theta$ is the potential
temperature, $\Theta_*$ is the so-called temperature scale, and
$c_{i,*}$ is the so-called scale of matter. Furthermore,
$\Phi_m(\zeta)$, $\Phi_h(\zeta)$, and $\Phi_i(\zeta)$ are the local
similarity functions for momentum (subscript m), sensible heat
(subscript h), and matter (subscript i), respectively. Moreover,
$\zeta$ is the Obukhov number, where $L$ is the Obukhov stability
length defined by \cite{Obuk, Monin, Zil, Kra89}
\begin{equation}\label{7.4}
L = \frac{u_*^2}{\kappa\;
\frac{g}{\overline{\Theta}}\;\bigl(\Theta_* + 0.61\;
\overline{\Theta}\; c_{v,*}\bigr)}\;.
\end{equation}
\noindent Here, $g$ is the acceleration of gravity, and $c_{v,*}$ is
the so-called humidity scale when subscript $i$ stands for water
vapor. The turbulent fluxes of momentum (the magnitude of the
Reynolds stress vector), $\tau$, sensible heat, $H$, and matter,
$F_i$, are related to these scaling quantities by \cite{Monin,
Busch, Pano84, Kra89}
\begin{equation}\label{7.5}
\tau = \Bigl(\overline{\rho}\;\overline{u'\;w'}+
\overline{\rho}\;\overline{v'\;w'}\Bigr)^{\frac{1}{2}} =
\overline{\rho}\;u_*^2 = const.\;,
\end{equation}
\begin{equation}\label{7.6}
H = c_{p,d}\;\overline{\rho}\;\overline{w'\;\Theta'} = -
c_{p,d}\;\overline{\rho}\;u_*\;\Theta_* = const.\;,
\end{equation}
\noindent and
\begin{equation}\label{7.7}
F_i = \overline{\rho}\;\overline{w'\;c_i'} = -
\overline{\rho}\;u_*\;c_{i,*} = const.\;,
\end{equation}
\noindent where $c_{p,d}$ is the specific heat at constant pressure
for dry air. Assuming that these fluxes are invariant with height
and integrating Eqs. (\ref{7.1}) to (\ref{7.3}) over he height
interval $[z_1, z_2]$ ($z_1$ and $z_2$ may be the lower and upper
boundaries of the fully turbulent part of the ASL) result in
\begin{equation}\label{7.8}
U(z_2) = U(z_1) + \frac{u_*}{\kappa}\Bigl(\ln\frac{z_2}{z_1}-
\Psi_m\bigl(\zeta_2,\zeta_1\bigr)\Bigr)\;,
\end{equation}
\begin{equation}\label{7.9}
\overline{\Theta}(z_2) = \overline{\Theta}(z_1) +
\frac{\Theta_*}{\kappa}\Bigl(\ln\frac{z_2}{z_1}-
\Psi_h\bigl(\zeta_2,\zeta_1\bigr)\Bigr)\;,
\end{equation}
\noindent and
\begin{equation}\label{7.10}
\overline{c_i}(z_2) = \overline{c_i}(z_1) +
\frac{c_{i,*}}{\kappa}\Bigl(\ln\frac{z_2}{z_1}-
\Psi_i\bigl(\zeta_2,\zeta_1\bigr)\Bigr)\;,
\end{equation}
\noindent where the corresponding integral similarity functions are
defined by \cite{Pano63, Pau, Pano84, Kra89}
\begin{equation}\label{7.11}
\Psi_{m,h,i}\bigl(\zeta_2,\zeta_1\bigr) = \int_{\zeta_1}^{\zeta_2}\frac{1 -
\Phi_{m,h,i}(\zeta)}{\zeta}\;d\zeta
\end{equation}
\noindent Obviously, not only the similarity hypotheses of Monin and
Obukhov but also the integration of Eqs. (\ref{7.1}) to (\ref{7.3}
require that the fluxes of momentum, sensible heat and matter are
height-invariant across the ASL. Since the local similarity
functions of Monin and Obukhov cannot be derived using methods of
dimensional analysis like Buckingham's \cite{Buck} $\pi$ theorem, it
is indispensable to use formulae empirically derived. Integral
similarity functions that are based on various local similarity
functions empirically determined are reviewed by Panofsky and
Dutton \cite{Pano84}, Kramm \cite{Kra04}, and Kramm and Herbert \cite{Kra08}.\\
\indent The set of so-called profile functions (\ref{7.8}) to
(\ref{7.11}) enables the experimentalist to estimate turbulent
fluxes of momentum, sensible heat and matter in dependence on the
thermal stratification of air at a certain height above the lower
boundary, namely the earth's surface, using the vertical profiles of
mean values of windspeed, potential temperature and mass fractions
\cite{Gar, Stu}. Furthermore, the parameterization schemes used in
state-of-the-art numerical models of the atmosphere (weather
prediction models, climate models, etc.) for predicting the exchange
of momentum, sensible heat and matter between the atmosphere and the
underlying ground also based on this set of profile functions.\\
\indent In the case of trace species Eq. (\ref{5.1}) is the
essential rule for micrometeorological profile measurement. With
respect to this, it requires stationary state and horizontally
homogeneous conditions and that chemical reactions play no role like
in the case of $CO_2$. Under these premises one obtains
\begin{equation}\label{7.12}
\frac{\partial}{\partial{z}}\Bigl(\overline{\rho_i'\;w'} +
\overline{J_{i,z}}\Bigr)= 0\Rightarrow \overline{\rho_i'\;w'} +
\overline{J_{i,z}} = \textrm{const.}
\end{equation}
\noindent This result completely agrees with Eq. (\ref{6.3}). On the
contrary, under the same premises the "alternative" equation
(\ref{5.4}) leads to
\begin{equation}\label{7.13}
\frac{\partial}{\partial{z}}\Bigl(\overline{\rho_i'\;w'}\Bigr)=
\overline{S_i}\;\delta \bigl(z\bigr) \neq \textrm{const.}
\end{equation}
\noindent when we assume for a moment that Eq. (\ref{6.9}) would be
correct. Consequently, the "alternative" advection-diffusion
equation (\ref{5.4}) would imply that no basis for the similarity
laws of Monin and Obukhov does exist, i.e., the profile functions
(\ref{7.8}) to (\ref{7.11}) customarily used for determining the
fluxes of long-lived trace gases like $CO_2$ would be obsolete.
\begin{figure}[t]
\begin{center}
\includegraphics[width=0.8\textwidth,height=!]{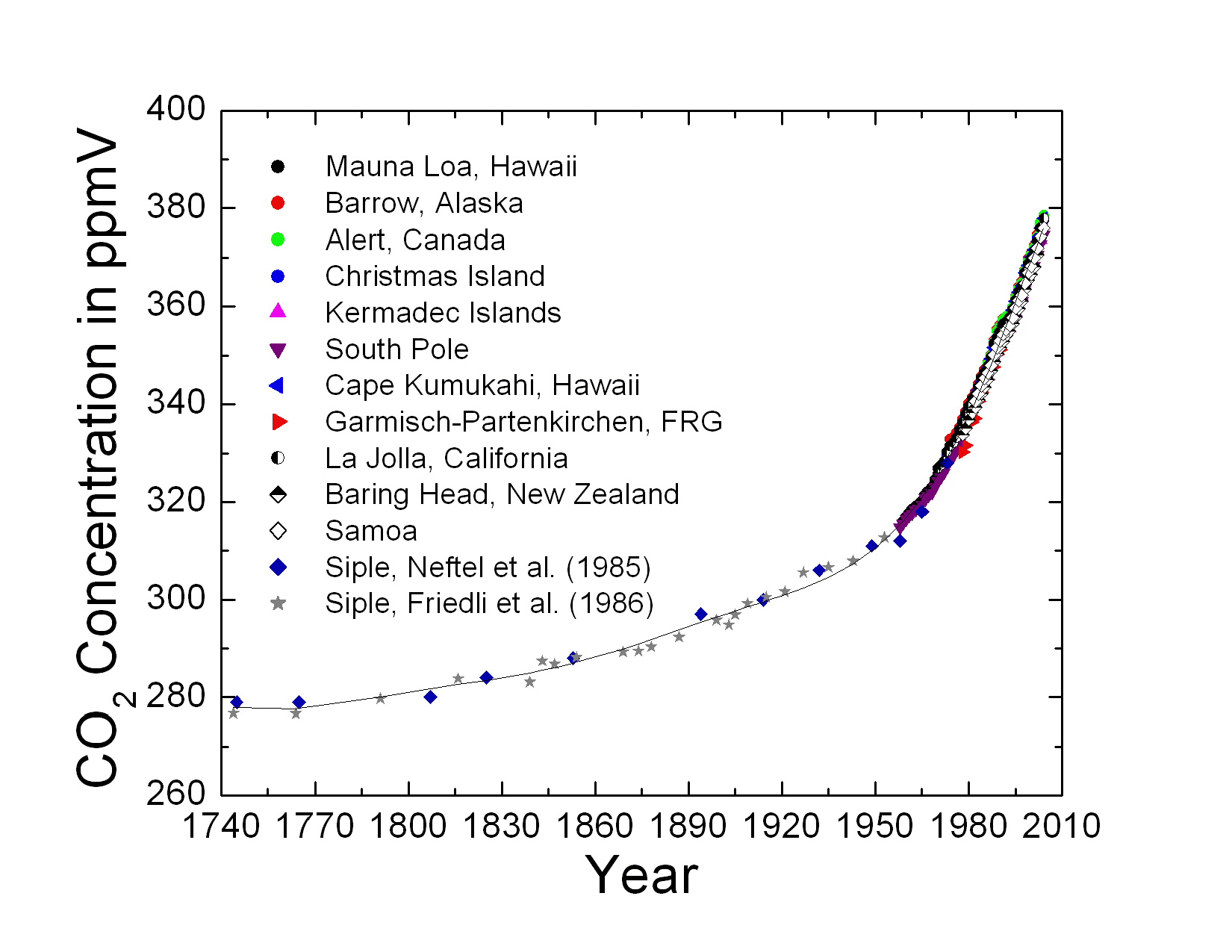}
\end{center}
\caption{Atmospheric $CO_2$ concentration inferred from Siple
station (West Antarctica) ice cores analyzed by Neftel et al.
\cite{Neft} and Friedli et al. \cite{Fried} and extended to that
directly observed at various locations during the period 1958-2004
(courtesy of C.D. Keeling, T.P. Whorf, and the Carbon Dioxide
Research Group of the Scripps Institution of Oceanography (SIO). The
fit is based of a polynomial of $8^{th}$ order.} \label{Figure4}
\end{figure}
\section{The global budget of carbon dioxide}
\noindent Dividing Eq. (\ref{2.7}) by the volume and rearranging the
resulting equation yield for a turbulent system
\begin{equation}\label{8.1}
\Bigl\langle\frac{\partial{\overline{\rho_i}}}{\partial{t}}\Bigr\rangle_V
=
-\;\frac{1}{V(t)}\;\int_{A(V(t))}\bigl(\overline{\rho_i}\;\overline{\textbf{v}}
+\overline{\rho_i'\;\textbf{v}'}+ \overline{\textbf{J}_i}\bigr)\cdot
d\textbf{A} + \bigl\langle\overline{\sigma_i}\bigr\rangle_V\;,
\end{equation}
\noindent where the volume average of an arbitrary quantity $\psi$
is generally defined by
\begin{equation}\label{8.2}
\bigl\langle\psi\bigr\rangle_V = \frac{1}{V}\int_V \psi \; V\;.
\end{equation}
\indent Hitherto, the volume $V(t)$  has been considered as
arbitrary. Suppose the earth can be considered as a sphere with the
radius $r_E \cong 6,371$ km and the atmospheric layer under study
that directly covers the earth as a spherical shell of a thickness
of $\Delta r$, the volume of the atmospheric layer, $V_A$, now
considered as independent of time is then given by
\begin{equation}\label{8.3}
V_A = \frac{4}{3}\;\pi\;r_E^3\;\Bigl\{\Bigl(1 + \frac{\Delta
r}{r_E}\Bigr)^3 - 1\Bigr\}\;.
\end{equation}
\noindent Since $\Delta r/r_E \ll 1$, the term $(1 + \Delta
r/r_E)^3$ can be approximated by $1 + 3\;\Delta r/r_E$. Thus, we
obtain 
\begin{equation}\label{8.4}
V_A = 4\;\pi\;r_E^2\;\Delta r\;.
\end{equation}
\noindent The surface of this spherical shell is given by $A_A = A_T
+ A_E = 4\;\pi\;r_E^2\;\{(1 + \Delta r/r_E)^2 + 2\}$, where $A_T$
and $A_E$ are the outer surface and the inner surface of this shell,
respectively. The latter is congruent with the earth's surface. The
surface integral in Eq. (\ref{8.1}) may, therefore, be expressed by
\begin{eqnarray}\label{8.5}
\nonumber
\lefteqn{\int_{A_A}\bigl(\overline{\rho_i}\;\overline{\textbf{v}}
+\overline{\rho_i'\;\textbf{v}'}+ \overline{\textbf{J}_i}\bigr)\cdot
d\textbf{A} =
\int_{A_T}\bigl(\overline{\rho_i}\;\overline{\textbf{v}}
+\overline{\rho_i'\;\textbf{v}'}+ \overline{\textbf{J}_i}\bigr)\cdot
d\textbf{A}} \\
&+& \int_{A_E}\bigl(\overline{\rho_i}\;\overline{\textbf{v}}
+\overline{\rho_i'\;\textbf{v}'}+ \overline{\textbf{J}_i}\bigr)\cdot
d\textbf{A}\;.
\end{eqnarray}
\noindent The first integral on the right-hand side of this equation
describes the exchange of matter between the atmospheric layer under
study and the atmospheric region aloft across the common border, and
the second integral the exchange of matter between the atmospheric
layer under study and the earth caused by the total (anthropogenic
plus natural) emission (counted positive) and the uptake (counted
negative) by the terrestrial biosphere (plants and soils) and the
oceans. If we choose $\Delta r$ in such a sense that exchange across
the outer surface of this spherical shell is equal to zero (or in
comparison with that at the earth's surface, at least, negligible)
because of the vertical profile of the $CO_2$ concentration becomes
(nearly) independent of height (see, e.g., Figure 1 in
\cite{Steph}), we will obtain
\begin{equation}\label{8.6}
\int_{A_A}\bigl(\overline{\rho_i}\;\overline{\textbf{v}}
+\overline{\rho_i'\;\textbf{v}'}+ \overline{\textbf{J}_i}\bigr)\cdot
d\textbf{A} =
\int_{A_E}\bigl(\overline{\rho_i}\;\overline{\textbf{v}}
+\overline{\rho_i'\;\textbf{v}'}+ \overline{\textbf{J}_i}\bigr)\cdot
d\textbf{A}\;.
\end{equation}
\noindent Since the unit vector normal to the inner surface of the
spherical shell shows in the direction of the earth's center, we
have only to consider the radial components of these fluxes
characterized by the subscript $r$, i.e.,
\begin{eqnarray}\label{8.7}
\nonumber \lefteqn{
\int_{A_E}\bigl(\overline{\rho_i}\;\overline{\textbf{v}}
+\overline{\rho_i'\;\textbf{v}'}+
\overline{\textbf{J}_i}\,\bigr)\cdot d\textbf{A} =
-\;\int_{A_E}\bigl(\overline{\rho_i}\;\overline{v_r}
+\overline{\rho_i'\;v_r'}+ \overline{J_{r,i}}\,\bigr)\Bigr|_{EM}
dA }\\
&+& \int_{A_E}\bigl(\overline{\rho_i}\;\overline{v_r}
+\overline{\rho_i'\;v_r'}+ \overline{J_{r,i}}\,\bigr)\Bigr|_U dA\;.
\end{eqnarray}
\noindent The first integral on the right-hand side of this equation
represents the total emission (subscript $EM$) and the second one
the uptake (subscript $U$). The signs of these terms are determined
by the scalar product between the unit vectors of the fluxes and the
unit vector normal to the inner surface of the spherical shell.\\
\indent Combining Eqs. (\ref{8.1}) and (\ref{8.5}) to (\ref{8.7})
yields 
\begin{eqnarray}\label{8.8}
\nonumber
\Bigl\langle\frac{\partial{\overline{\rho_i}}}{\partial{t}}\Bigr\rangle_{V_A}
&=&
\frac{1}{V_A}\;\Bigl\{\int_{A_E}\bigl(\overline{\rho_i}\;\overline{v_r}
+\overline{\rho_i'\;v_r'}+ \overline{J_{r,i}}\,\bigr)\Bigr|_{EM}dA \\
&-& \int_{A_E}\bigl(\overline{\rho_i}\;\overline{v_r}
+\overline{\rho_i'\;v_r'}+ \overline{J_{r,i}}\,\bigr)\Bigr|_U
dA\Bigr\} + \bigl\langle\overline{\sigma_i}\bigr\rangle_{V_A}\;.
\end{eqnarray}
\noindent Since $CO_2$ is a long-lived trace gas, the effect caused
by chemical reactions can be ignored. Thus, Eq. (\ref{8.8}) can be
approximated by
\begin{eqnarray}\label{8.9}
\nonumber
\Bigl\langle\frac{\partial{\overline{\rho_i}}}{\partial{t}}\Bigr\rangle_{V_A}
&=&
\frac{1}{V_A}\;\Bigl\{\int_{A_E}\bigl(\overline{\rho_i}\;\overline{v_r}
+\overline{\rho_i'\;v_r'}+ \overline{J_{r,i}}\,\bigr)\Bigr|_{EM}dA \\
\nonumber &-& \int_{A_E}\bigl(\overline{\rho_i}\;\overline{v_r}
+\overline{\rho_i'\;v_r'}+ \overline{J_{r,i}}\,\bigr)\Bigr|_U dA\Bigr\}\;.\\
\end{eqnarray}
\noindent Expanding the right-hand side of this
equation with $A_E$ yields finally
\begin{eqnarray}\label{8.10}
\nonumber
\Bigl\langle\frac{\partial{\overline{\rho_i}}}{\partial{t}}\Bigr\rangle_{V_A}
&=& \frac{1}{\Delta r}
\;\Bigl\{\Bigl\langle\bigl(\overline{\rho_i}\;\overline{v_r}
+\overline{\rho_i'\;v_r'}+ \overline{J_{r,i}}\,\bigr)\Bigr|_{EM}\Bigr\rangle_{A_E} \\
&-& \Bigl\langle\bigl(\overline{\rho_i}\;\overline{v_r}
+\overline{\rho_i'\;v_r'}+ \overline{J_{r,i}}\,\bigr)\Bigr|_U
\Bigr\rangle_{A_E} \Bigr\}\;,
\end{eqnarray}
\noindent where
\begin{equation}\label{8.11}
\bigl\langle\psi\bigr\rangle_{A_E} =
\frac{1}{A_E}\int_{A_E}\psi\;dA\;.
\end{equation}
\noindent represents the global average of an arbitrary quantity
$\psi$. Equation (\ref{8.10}) deduced by using the exact integral
formulations substantiates that (a) total emission and the uptake of
$CO_2$ have to be considered as lower boundary conditions and (b)
the partial density of $CO_2$ averaged over the volume of the
atmospheric layer under study will rise as long as the globally
averaged total emission is higher than the globally averaged uptake.
\section{Final remarks and conclusions}
\noindent  In 1999 Sarmiento and Gruber \cite{Sar} pointed out that
"the land sink for carbon is the subject of considerable controversy
at present, concerning not only its magnitude but also its cause".
It seems that any use of an "alternative" mass balance equation in
micrometeorology may contribute to considerably more confusion
because this "alternative" expression is clearly incorrect.\\
\indent The use of an "alternative" mass balance equation can
harmfully affect not only the whole atmospheric budget of $CO_2$,
but also that of other greenhouse gases like water vapor (the most
important greenhouse gas), nitrous oxide ($N_2O$), methane ($CH_4$),
and ozone ($O_3$).\\
\indent If Eqs. (\ref{3.1}) and (\ref{5.4}) are correct, the
biological source/sink, for instance, would explicitly cause a
temporal change in the partial density. However, the reality is
different. Figure \ref{Figure4} illustrates that the atmospheric
$CO_2$ concentration has been rising since the beginning of the 18th
century. If these results are correct as stated in various reports
of the Intergovernmental Panel on Climate Change (IPCC), we can
infer from Eq. (\ref{8.10}) that the total emission of $CO_2$ has
always been higher than the $CO_2$ uptake during the period covered
by these results. Thus, lowering, for instance, the anthropogenic
emissions of $CO_2$ to those of the year 1990 would not reduce the
$CO_2$ concentration in the atmosphere (see Special Report on
Emissions Scenarios (SRES) of the Working Group III of the
Intergovernmental Panel on Climate Change (IPCC) and the IPCC IS92,
\cite{Hou, Kat}). However, if the $CO_2$ uptake would rise due to a
higher atmospheric $CO_2$ concentration, a stabilization of this
concentration at a level of nearly 550 ppmV, i.e., higher than
that of 1990, might be possible \cite{Hou, Kat}.\\
\indent A decrease of the atmospheric $CO_2$ concentration can only
be achieved when the $CO_2$ uptake by the terrestrial biosphere and
the ocean is higher than the total emission $CO_2$. As indicated by
Beck's\cite{Beck} inventory for the past 180 years that is based on
more than 90,000 observations using chemical methods, there were
$CO_2$ concentrations appreciably higher than the current value of
about 385 $ppmV$. Only when Beck's data are reliable, we may
conclude that during various periods of the past the uptake was
stronger than the total emission.
\section*{Acknowledgments} Sections 2 to 6 of this contribution
were presented and discussed on the workshop of the University of
Alaska Fairbanks "2007 Dynamics of Complex Systems: Common Threads"
held in Fairbanks, Alaska, July 25-27, 2007. We would like to
express our thanks to the conveners of this workshop.

\end{document}